\documentclass[12pt,a4paper]{article}
\usepackage[dvips]{graphicx}
\begin{document}

\def\br{{\bf r}}
\def\bQ{{\bf Q}}

\title{\bf Effects of droplet fluctuations on the scattering
of neutrons and light by microemulsions}

\author{V.~Lis\'y and B.~Brutovsk\'y\\Department of Biophysics, P. J. \v
Saf\'arik University\\Jesenn\'a 5, 04154 Ko\v sice, Slovakia\\e-mail:
lisy@kosice.upjs.sk}

\maketitle

\begin{abstract}

Beginning from the first neutron spin-echo study of the shape
fluctuations of microemulsion droplets [J. S. Huang, S. T. Milner,
B. Farago, and D. Richter, Phys. Rev. Lett. {\bf 59}, 2600 (1987)]
these experiments are incorrectly interpreted in the literature.
This is due to an inappropriate account for the fluctuations and the
erroneous application of the original theory to the description
of the experiments (see [V. Lisy and B. Brutovsky, Czech J. Phys. {\bf 50},
239 (2000)]). In the presented work both these shortcomings are
corrected. We develop the theory of static and dynamic light and 
neutron scattering from droplet microemulsions. 
The fluids inside and out of the droplets are separated by a surfactant layer of 
arbitrary thickness. The scattering functions consistently take into account thermal fluctuations 
of the shapes of such double-layered spheres to the second order in the changes of their radius. 
The relaxation times and correlation functions of the fluctuations are found within the 
Helfrich's theory of interfacial elasticity. The theory is applied to the quantitative description 
of small-angle neutron scattering, neutron spin-echo spectroscopy and dynamic light scattering 
experiments. Basic characteristics of the microemulsions, extracted from the fits to the 
experimental data, significantly differ from those determined in the original works. We include 
into the consideration the viscosity of the surface layer and give its estimation for 
the octane - C$_{10}$E$_5$ - water microemulsion.

\end{abstract}

\section*{Introduction}

   Microemulsions, thermodynamically stable homogeneous dispersions of oil 
and water, are intensively studied because of their interesting physical
properties and practical importance \cite{Shah1}. The properties of 
the microemulsions are to a large degree determined by the characteristics
of the interface between the bulk fluids. Within the commonly accepted 
Helfrich's phenomenology \cite{Helfrich1}, this surface layer is described
by a few basic parameters. For a microemulsion droplet these are the bending
and Gaussian moduli $\kappa$ and $\overline{\kappa}$, the spontaneous
curvature $C_s$, the surface tension coefficient $\alpha$, and the equilibrium
radius of the droplet, $R_0$. The determination of these characteristics
has been attempted by several methods \cite{Hellweg1}. However, different
approaches yield very different values of the parameters of the surface film.
So, the neutron spin-echo (NSE) technique of the quasielastic neutron
scattering gives significantly larger values $\kappa$ than indirect
macroscopical or optical techniques \cite{Hellweg1}. This coefficient is 
in the NSE experiments obtained from the analysis of the peak observed
in the $Q$ (wave-vector transfer at the scattering) dependence 
of the effective diffusion coefficient of the droplets, $D_{eff}(Q)$.
The characteristics of the peak are connected with the thermal shape 
fluctuations of the droplets \cite{Hellweg1,Farago1}.
In the present work the role of these fluctuations is carefully studied 
and it is shown that their influence on the observed quantities was
underestimated in the interpretation of the scattering experiments.
We calculate the intermediate scattering function $F(Q,t)$ that
is directly measured by NSE and suitable for the description 
of small-angle neutron scattering (SANS) and dynamic scattering
of light (DLS) as well. As distinct from the previous works,
$F(Q,t)$ is found taking into account all the contributions up to the 
second order of the fluctuations of the droplet radius.

\section*{The scattering functions}

The microemulsion droplet is modeled by a double-layered sphere
with the surface layer of a constant thickness $d=R_2-R_1$, 
where $R_1=R_0-d/2$ is its inner and $R_2$ outer radius. Inelastic 
scattering of neutrons is determined by the time evolution of the 
scattering length density $\rho(\br,t)$ \cite{Lovesey1} (in the case of DLS, 
$\rho(\br,t)$ is replaced by the dielectric constant 
${\epsilon(\br,t)}$ \cite{Komarov1}).
It is assumed that the scattering from individual droplets is essential. 
Let the scattering length density is $\rho_1$ in the droplet interior, 
$\rho_2$ in its exterior, and the surface shell is characterized by the 
constant $\rho_0$. The deviation from the radius $R_0$ of the nondeformed 
shell (which is also, within the approximation used below, 
the equivalent-volume radius) is described by the quantity $u=R(t)-R_0$.
The quasielastic scattering of neutrons in Born approximation is given
by the Van Hove function for coherent scattering, 
$F(Q,t)=\langle\rho(\bQ,t)\rho^{\ast}(\bQ,0)\rangle$. To calculate this correlator, 
$u/R_0$ is expanded in spherical harmonics \cite{Milner1}, with the coefficients
$u_{lm}(t)$, where $m=-l, -l+1, \ldots , l,$ and $l=0, 1, \ldots , l_{max}
\sim R_0/a$ ($a$ is a typical molecular diameter). 
The fluctuations of the molecules 
in the layer are neglected since the layer is usually thought to be almost
incompressible \cite{Milner1}. The shape fluctuations with different numbers $l$ and 
$m$ are uncorrelated and $\langle u_{lm}\rangle = 0$ for $l>1$. Using the known properties 
of the spherical harmonics and spherical Bessel functions $j_l$, we obtain 
to the second order in the fluctuations $(Q\ne0)$

\vspace{5mm}

\begin{equation}
\label{Principal}
\frac{F(Q,t)}{(4 \pi R_0^2 d \Delta)^2} =
\Phi^2+2\Phi\Phi_0\frac{\langle u_{00}\rangle}{\sqrt{4\pi}}+\sum_{l>1}
\frac{2l+1}{4\pi}
\bigg\{\Phi\Psi\langle u_{l0}^2\rangle+\Phi_l^2\langle u_{l0}(0)u_{l0}(t)
\rangle\bigg\},
\end{equation}

$$
\Phi(Q)=\frac{\rho(Q)}{4\pi R_0^2 d \Delta}, \quad
\rho(Q)=\int e^{iQr}\rho({\br})d\br=4\pi\bigg\{(\rho_1-\rho_0)R_1^3
\frac{j_1(x_1)}{x_1}+(\rho_0-\rho_2)R_2^3\frac{j_1(x_2)}{x_2}\bigg\},
$$

$$
\Phi_l(Q)=\frac{1}{R_0d\Delta}
\Bigl[(\rho_1-\rho_0)R_1^2 j_l(x_1)+(\rho_0-\rho_2)R_2^2 j_l(x_2)\Bigr],
$$

$$
\Psi(Q)=\frac{1}{d\Delta}
\Bigl[(\rho_1-\rho_0)R_1 \varphi(x_1)+(\rho_0-\rho_2)R_2 \varphi(x_2)\Bigr], 
\quad
\varphi(x)=2j_0(x)-xj_1(x),
$$

\vspace{5mm}

\noindent where $x_{1,2}=QR_{1,2}$, and $\Delta=(\rho_1+\rho_2)/2-\rho_0$ characterizes 
the contrast between the densities of the bulk fluids and the surface shell.
The $l=1$ mode automatically dropped out during the calculations. 
This coincides with the fact known for the droplets with thin adsorbed 
layers \cite{Milner1} that this mode corresponds to their translational motion. 
For $l=1$ there is no motion in the layer and the droplet moves as 
a hard sphere \cite{Lisy1,Lisy2}. The calculation was carried out 
in the system connected with the droplet;
as usually, the effect of translational motion 
is included assuming the statistical independence of the translational 
and other degrees of freedom. Then the right-hand side of eq. 
(\ref{Principal}) has to be multiplied by the factor exp$(-Q^2Dt)$,
where $D$ is the self-diffusion coefficient of a sphere with 
the hydrodynamic radius $R_H=R_2+\delta$ ($\delta$ accounts 
for the shell of solvent molecules that move together with the droplet).

Equation (\ref{Principal}) is the basic result of the paper. The first term 
in its right side determines the formfactor of the scattering in the absence 
of fluctuations \cite{Guinier1}. The next terms are connected with the shape fluctuations.
For the first time all the second-order terms in the fluctuating density 
(or dielectric constant) expansion are taken into account. Due to this 
eq. (\ref{Principal}) significantly differs from the expressions for the function
$F(Q,t)$ known from the literature. The main difference is in the second term that
is given by the $l=0$ mode contribution. This term did not appear in previous works,
though for incompressible bulk fluids
$\langle u_{00}\rangle = -\Sigma(2l+1)\langle u_{l0}u_{l0}\rangle
/\sqrt{4\pi}$ $(l>1)$ is nonzero \cite{Milner1}.
Also the first term in the sum in eq. (\ref{Principal}) usually absents
in the description of the scattering experiments \cite{Hellweg1}.
The correct expression
for this contribution was found by Gradzielski et al. \cite{Gradzielski1}
for the conditions of the perfect shell contrast (deuterated bulk
fluids with the densities
$\rho_1\approx\rho_2$ significantly different from the density of the hydrogenated
layer). In all the works the scattering function was calculated for $d$ small
in comparison with the droplet radius and does not contain the terms
$\sim (\rho_1-\rho_2)$. 
It is thus inappropriate for the description of the scattering in conditions 
far from the shell contrast. In the $d/R_0\to 0$ limit the previous expressions
for $F(Q,t)$ converge to 0 so they cannot describe the scattering from large 
emulsion or vesicle droplets. The last term in eq. (\ref{Principal}) agrees
with that from the literature for small $d/R_0$ and $\rho_1=\rho_2$.

Equation (\ref{Principal}) must be completed by expressions for the correlation
functions of the fluctuating droplet radius. In the harmonic approximation 
for isothermal fluctuations, small thicknesses of the surface layer, 
and the conservation of the droplet volume as well as the total number 
of molecules in the shell, we have \cite{Lisy3}

\begin{equation}
\label{CorrFunction}
\langle u_{l0}(t)u_{l0}(0) \rangle \approx \frac{k_BT}{\alpha_lR_0^2(l+2)(l-1)}
exp(-\omega_lt), \ \alpha_l=\alpha-2\kappa C_sR_0^{-1}+\kappa l(l+1)R_0^{-1}.
\end{equation}

\vspace{5mm}

\noindent
The decay rates of the fluctuations, $\omega_l$, have been studied in a number 
of papers \cite{Lisy1,Lisy2}. According to Fujitani \cite{Fujitani1}, the relaxation modes of incompressible 
dissipative layers are described by the formula

\begin{equation}
\label{Fujitani}
\omega_l = \frac{\alpha_l}{R_0} \ 
\frac{l(l+1)(l-1)(l+2)}{4(l-1)(l+2)\eta_0/R_0 + (l-1)(2l^2+5l+5)\eta_1
+ (l+2)(2l^2-l+2)\eta_2},
\end{equation}

\vspace{5mm}

\noindent
where $\eta_{1,2}$ are the viscosity coefficients of the bulk fluids and $\eta_0$
is the viscosity of the layer. Finally, the experimental samples contain droplets 
with different radii so that the measured quantities are the averages over
the distribution of the droplets in radii. Within the phenomenological theory 
of the droplet formation, the distribution function $f(R_0)$ has a form 
$f(R_0)\sim exp[-(1-R_0/R_m)^2/2\epsilon]$, where $R_m$ is the mean radius 
of the droplets, $\langle (R_0-R_m)^2\rangle \approx \epsilon R_m^2$, 
and $\epsilon$ is the polydispersity functionally related to the characteristics
of the layer \cite{Lisy3}. In the simplest case of the so-called two-phase coexistence 
$\epsilon=k_BT/8\pi(2\kappa+\overline{\kappa})$. Note that the entropy-of-mixing
contribution is easily incorporated in this relation by adding $2k_BTF(\phi)$
to the denominator \cite{Gradzielski1}. In the random mixing approximation, for small droplet 
volume fractions $\phi$, the function $F(\phi)\approx ln\phi-1$. Analogously, 
eqs. (\ref{CorrFunction}) and (\ref{Fujitani}) can be appropriately changed 
by replacing $R_0^2\alpha_l\to\kappa(l-1)(l+2)-\overline{\kappa}-k_BTF(\phi)/4\pi$.
This allows us to express the mean quadrate of the fluctuations in the form

\begin{equation}
\label{QuadForm}
\langle u_{l0}^2\rangle =
\Bigg\{(l-1)(l+2)\biggl[\frac{\kappa}{k_BT}l(l+1)-\frac{1}{8\pi\epsilon}
\biggr]\Bigg\}^{-1}
\end{equation}

\vspace{5mm}

\noindent
that is particularly suitable to be used in fits to experimental data since it is 
invariant with respect to the function $F(\phi)$.

The analysis of eqs. (\ref{Principal} - \ref{Fujitani}) shows that for a number
of microemulsion systems studied in the literature the effect of the fluctuations
on the scattering functions is important. The fluctuations can significantly affect
the position of the minimum in the $Q$ dependence of the scattering intensity measured
by SANS, the position of the peak in $D_{eff}(Q)$ observed by NSE, as well as
the diffusion coefficient determined using DLS. As a result some basic 
characteristics of the microemulsions, extracted by us from these experiments,
notably differ from those found in the works where the fluctuations were ignored
(SANS) \cite{Hellweg1,Gradzielski1} or where the account for them should
be corrected (NSE, DLS) \cite{Hellweg1,Farago1}.

\section*{Quantitative description of the scattering experiments}

\begin{flushright}
\begin{figure}[!h]

\begin{center}
\includegraphics{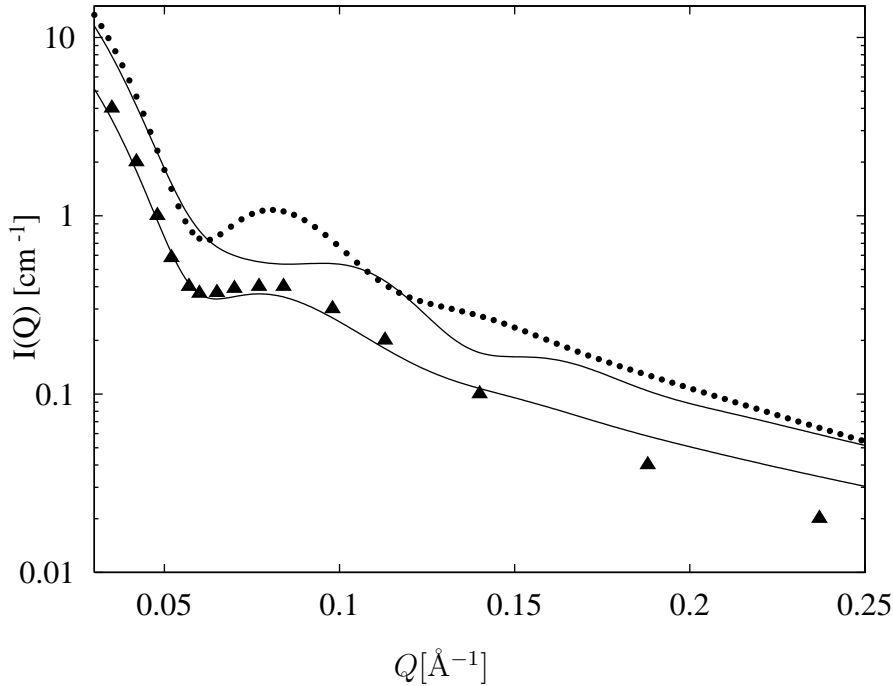}

$Q$[\AA$^{-1}$]

\end{center}

\caption[I(Q)]
{\footnotesize The SANS intensity calculated from eqs. 
(\ref{Principal}-\ref{CorrFunction}) with (the upper full line)
and without (dashed line) the droplet fluctuations in the shape.
The experimental points are from the work \cite{Hellweg1}.
The octane-$d_18$ - surfactant (C$_{10}$E$_5$) - water microemulsion 
at the temperature $T=305.2$ K is characterized by the parameters
$\rho_1=6.35 \times 10^{10}$, $\rho_2=6.36 \times 10^{10}$ and 
$\rho_0=1.65 \times 10^9 cm^{-2}$, $\kappa=0.92 k_BT$, $\epsilon=0.036$,
$d=10.8$\AA, $R_m=48$\AA, and the volume fraction of the droplets 
is $0.049$ \cite{Hellweg1}. Our fit to the experimental data
yielded the values $\kappa=1.93 k_BT$, $\epsilon=0.046$,
$R_m=47.7$\AA\ and $d=6.4$\AA.}
\label{IQ}
\end{figure}
\end{flushright}

We applied the obtained formulae to describe several light and neutron
scattering experiments. In the case of large emulsion droplets (the limit $d\to 0$)
eq. (\ref{Principal}) well describes the light diffusing-wave spectroscopy 
experiments \cite{Lisy4,Lisy5}. Here we present 
the results of the description of the experiments
on a microemulsion system \cite{Hellweg1} that well illustrate
the importance of the correct account for the effects of fluctuations.
In the work \cite{Hellweg1} the octane-$d_{18}$ - surfactant
(C$_{10}$E$_5$) - water microemulsion 
in the conditions of two-phase coexistence was studied using SANS, NSE and DLS.
The SANS experiments were interpreted using the model of nonfluctuating 
double-layered spheres \cite{Guinier1}. For a diluted microemulsion dispersion the interaction
between the droplets was negligible in the region of $Q>0.03$\AA$^{-1}$ so that
the static structure factor $S(Q)\approx 1$. We have thus calculated the SANS
intensity according to the formula $I(Q)\approx N<\!\!F(Q,0)\!\!>$
\cite{Hellweg1}, where $N$ 
is the number density of the droplets and the average is over the distribution
$f(R_0)$. Figure \ref{IQ} shows the intensity $I(Q)$ calculated in the case of
nonfluctuating droplets and for fluctuating droplets characterized by 
the parameters determined in \cite{Hellweg1}. It is seen that both the curves 
do not describe the experimental data. From our fit to the experiment using
the scattering length densities $\rho_1=6.35\times 10^{10}$,
$\rho_2=6.36\times 10^{10}$ \cite{Gradzielski1} and 
$\rho_0=1.65\times 10^9 cm^{-2}$ \cite{Langevin1},we obtained
the set of the parameters $R_m$, $d$, $\epsilon$ 
and $\kappa$ for which the theoretical curve well reproduces the measured
$I(Q)$.

\begin{figure}[!h]

\begin{center}
\includegraphics{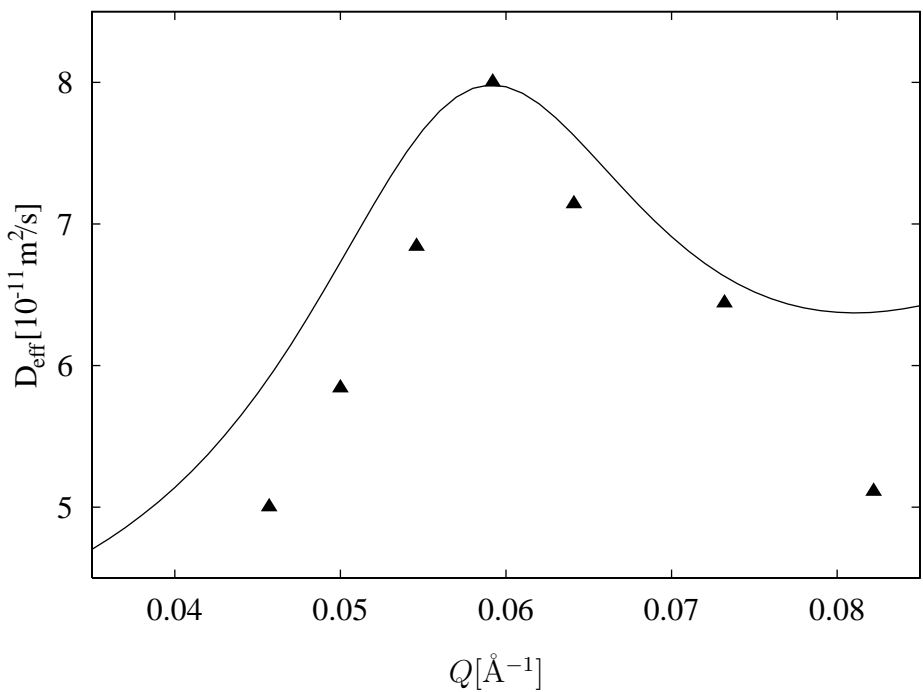}

$Q$[\AA$^{-1}$]

\end{center}

\caption[Deff]
{\footnotesize The effective diffusion coefficient measured by NSE 
(points) \cite{Hellweg1} on the same system as in Fig. \ref{IQ}, and calculated from 
eqs. (\ref{Principal}-\ref{QuadForm}) using the parameters obtained from the fit to SANS 
(Fig. \ref{IQ}). The height of the peak corresponds to the experimental
one if the viscosity of the surface layer $\eta_0=5.6 \times
10^{-11} Ns/m$ is employed in the calculations. The thickness
of the hydration shell $\delta=17$\AA\ was obtained from the DLS 
experiment \cite{Hellweg1}. The viscosities of the bulk fluids
were $\eta_1=0.4675$ and $\eta_2=0.7632$ cP.}
\label{Deff}
\end{figure}

\noindent
With these parameters and for dielectric constants $\epsilon_1=1.946$,
$\epsilon_2=1.769$ and $\epsilon_0=2.106$ \cite{Langevin1}, we evaluated the effective 
diffusion coefficient of the droplets, $D_{eff}=-<\!\!dF(Q,t)/dt\!\!>\!\!\!/
(Q^2\!\!<\!\!F(Q,0)\!\!>)$ at $t=0$.
By DLS this coefficient is measured in the region of $Q$ much smaller than
in the case of SANS. In this region $D_{eff}$ is practically constant, not influenced
by the time-dependent terms $\sim\langle u(0)u(t)\rangle$ in eq. (\ref{Principal}).
The Stokes self-diffusion coefficient has a form $D=k_BT/6\pi\eta_2R_H$,
$R_H=R_0+d/2+\delta$, and its mean value was determined from the requirement
to satisfy the measured $D_{eff}=3.8\times 10^{-11} m^2/s$ (with the tolerance
$\pm5\%$). 
Our attempt to describe the DLS experiment assuming a hydration shell with
a small $\delta$ (2-3 \AA) was not possible without a change of the parameters
$R_m$, $d$, $\epsilon$ and $\kappa$, that makes worse the agreement with the 
SANS experiment. The best agreement with both the experiments required 
$\delta=13 \div 21$ \AA (the mean $\delta\approx 17$ \AA).
The found thickness of the hydration shell assumes the shell built of several 
layers \cite{Hellweg1}. Such a possibility is a question for future investigations. We suppose 
however that first a model taking into account different properties of the bound 
and free bulk fluids should be developed. Finally, we turned to the NSE experiment.
This experiment was in the work \cite{Hellweg1} described
using the scattering function (1) without the time-independent
contributions of the fluctuations and for the ideal contrast.
The decay rates were from our paper \cite{Lisy6}, however,
those $\omega_l$ correspond
to infinitely easily compressible layers \cite{Lisy1,Lisy2}.
In our analysis we used
$\omega_l$ from eq. (\ref{Fujitani}) appropriate for incompressible layers.
The viscosities of the bulk fluids were $\eta_1=0.4675$ and $\eta_2=0.7632$ cP
at $T=305.9$ K \cite{Hellweg1}. The last parameter needed to describe the experiment is
$\eta_0$.
If the previous theories are correctly used the height of the peak
in $D_{eff}$ is several times larger than its measured values.
Also in our theory this height is too large. 
We propose to solve this problem 
introducing the dissipation in the layer that leads to a decrease of
$\omega_l$ and thus to a decrease of the peak height.
From the correspondence between the experimental \cite{Hellweg1} 
and theoretical heights 
of the peak we found $\eta_0=5.6\times 10^{-11}$ Ns/m, the value only 
weakly influenced by the parameter $\delta$. The calculated and experimental
$D_{eff}$ is shown on Figure \ref{Deff}. The rest of the parameters estimated from 
the fits are as follows: $\kappa=1.93 (0.92) k_BT$, $\epsilon=0.046 (0.036)$,
$\overline{\kappa} = -2.68 (-0.38) k_BT$, $R_m=47.7 (48)$\AA\ and $d=6.4
(10.8)$\AA\ (in parentheses the values from \cite{Hellweg1} are shown).
When the entropy-of-mixing
contributions are taken into account, only the parameter
$\overline{\kappa}$ slightly changes, now being $\overline{\kappa}=-3k_BT$.
The hydrodynamic radius of the droplets and the diffusion 
coefficient are about $R_H=68 (81)$\AA\ and $D(R_m)=4.3 (3.6) \times
10^{-11} m^2/s$.

\noindent
The difference between the obtained parameters and those found in the original
work \cite{Hellweg1} concerns first of all the elastic coefficients that are the crucial
parameters needed to describe the properties of microemulsions. For the first
time we have from the scattering experiments estimated the viscosity coefficient
of the surface layer.

In conclusion, a deeper consideration of the effects of the shape fluctuations
of microemulsion droplets on the static and dynamic neutron and light scattering
was presented. The agreement with such experiments is much better than it was
possible so far. More reliable information can be now obtained about important
phenomenological parameters of microemulsions. We suppose that in further
investigations the role of the hydration layer bound to the droplets should
be studied in detail. More should be also known about the interaction between
the droplets and the entropy of dispersion that are differently treated
in the literature \cite{Gradzielski1,Ruckenstein1}.

\vspace{1cm}

This work was supported by the grant No. 1/7401/20, VEGA, Slovak Republic.


\end{document}